# A Neural-Astrocytic Network Architecture: Astrocytic calcium waves modulate synchronous neuronal activity

Ioannis Polykretis
Rutgers University
Computational Brain Lab
New Brunswick, New Jersey

Vladimir Ivanov
Rutgers University
Computational Brain Lab
New Brunswick, New Jersey

Konstantinos P. Michmizos*
Rutgers University
Computational Brain Lab
New Brunswick, New Jersey

## ABSTRACT

Understanding the role of astrocytes in brain computation is a nascent challenge, promising immense rewards, in terms of new neurobiological knowledge that can be translated into artificial intelligence. In our ongoing effort to identify principles endowing the astrocyte with unique functions in brain computation, and translate them into neural-astrocytic networks (NANs), we propose a biophysically realistic model of an astrocyte that preserves the experimentally observed spatial allocation of its distinct subcellular compartments. We show how our model may encode, and modulate, the extent of synchronous neural activity via calcium waves that propagate intracellularly across the astrocytic compartments. This relationship between neural activity and astrocytic calcium waves has long been speculated but it is still lacking a mechanistic explanation. Our model suggests an astrocytic "calcium cascade" mechanism for neuronal synchronization, which may empower NANs by imposing periodic neural modulation known to reduce coding errors. By expanding our notions of information processing in astrocytes, our work aims to solidify a computational role for non-neuronal cells and incorporate them into artificial networks.

## CCS CONCEPTS

• **Computing methodologies** → **Model verification and validation**; • **Applied computing** → **Biological networks**; **Systems biology**; *Bioinformatics*;

## KEYWORDS

Neural Astrocytic Networks (NANs), Brain-morphic Architecture, Non-Von Neumann computing model, Event-based system

## INTRODUCTION

Innervated by Alan Turing's "heretic" theories[96], brain-morphic machines originate from the work of McCulloch and Pitts[53] who introduced the notion of binary state models of neurons that became the computational units in the first neural networks, namely multilayer perceptrons[74], Hopfield networks[40], and Boltzmann machines[37]. As Brain and Computing Sciences grow separately, strong criticisms, stemming from inconsistencies between neurobiology and neural networks[15], impeded the effective application of neuromorphism to real-world problems. It is Carver Mead's persistence on neuromorphic microelectronics[54] that ignited a number of large-scale neuromorphic systems, ranging from the Spaun software[22], a spiking version of a Restricted Boltzmann Machine, to hardware, namely Intel's Loihi[17] and IBM's TrueNorth chips[55], the Neurogrid[7] as well as the BrainScaleS and SpiNNaker [28], the last two being empowered by the EU's Human Brain Project. Targeting basic cognitive abilities, these neuro-morphic systems remain fairly restricted in hand-crafted applications as they currently offer rather limited options for brain plasticity mechanisms and lack any of the emerging principles that non-neuronal cells can bring into brain-inspired computing.

Spurred by the recent advancements in glial-specific Calcium imaging[98], a new interest in the electrically silent non-neuronal cells is starting to reveal their key computational roles in brain function[6, 29, 60]. Long believed to only give structural and nutritional support to neurons, astrocytes, the most well-studied and abundant type of glial cells, are now found to receive input from neurons and provide input to them, forming an additional computation layer[3, 6, 14, 16, 30, 45, 63, 70, 97]. Astrocytes differ from neurons in their range of neuromodulatory function as well as the time scale of doing so. Specifically, neurons transmit information fast via spikes, while astrocytes communicate communicate at a slow pace, through gap junctions with each other[14, 30] and through tripartite synapses with neurons[3, 16, 45, 63, 70, 97]. Interneurons also form meshes through gap junctions and exhibit calcium-dependent dynamics but release a single type of neurotransmitter, either excitatory [11] or inhibitory[51], whereas astrocytes can simultaneously excite or inhibit a synapse. Interestingly, this astrocytic-neural interaction is dynamic[38] and plastic[8, 93] and mediated mainly by waves of intracellular Calcium concentrations propagated at local[20], cellular[30, 38] and network scales[13, 99]. The roles of the astrocytic Calcium signals remain largely unknown[6, 98] because experimental results are limited to the cellular[101] or sub-cellular (synaptic)[20, 50] scale and, therefore, lack the spatial breadth required to study astrocytic function at the network level.

Being a functional part of the brain knot, astrocytes are now regarded as key cells affecting synaptic learning[68, 91] and neuronal oscillations[26], which are regarded as the functional building blocks for processing, transferring and learning information in the brain[35, 69]. Specifically, by virtue of their large spatial and temporal scales of interaction, their dynamics, their connectome, and their distinctive morphology, astrocytes are pivotal in synaptic depression, facilitation and pruning, as well as neurogenesis[24]. Furthermore, astrocytes are highly active during motor behaviors[62], and changes in astrocytes' physiology are found to directly impair fine motor coordination[76], which is consistent with the hypothesis that astrocytes are involved in sequence memory, as we also suggested recently by having astrocytes learn transition sequences on a Hopfield Neural-Astrocytic Network[46]. Astrocytes have also been experimentally linked to both slow-wave rhythmogenesis[100], as well as sleep pressure[57], which was further reinforced by another computational study of ours[47]. Interestingly, not only do astrocytes modulate changes in the strength of individual synapses along a neural network; they also form their own network[78], a global communication network overseeing and regulating critical brain functions that include the integration of neural information across spatial and temporal domains[61]. Not surprisingly, most brain diseases are now seen as conditions where both neurons and astrocytes are affected[4].

Therefore, developing comprehensive NANs that can explain high-level brain dynamics will illuminate the brain mechanisms underlying intelligent behavior, and hopefully facilitate new translative avenues between biological and artificial intelligence. The slower time scale in which astrocytes operate is well suited for analog operations in neuromorphic hardware, where they can be used to synchronize the neural spiking activity or, as we have shown, modulate it to transition between memorized states [46]. We envision that the development of comprehensive NANs, which can explain high-level brain dynamics can be embodied in neuromorphic hardware devices. The slower time scale in which they function might be well-suited for analog circuits. Although our understanding of neural-astrocytic interaction is still growing, translating this new "bottom-up" knowledge of astrocytic function into artificial intelligence is a timely and promising research direction. We have reached the point now where we can reasonably begin abstracting away from astrocyte biology and framing general computational principles. Although in its infancy, the application of glial biology to problems in artificial networks is on the rise, and has seen some early and encouraging successes[71].

Here, we propose a bio-mimetic astrocytic model that can be seamlessly integrated into one of our previously developed NAN architectures[46, 47] and process information in two different temporal scales, seconds and milliseconds. We show that the slow Calcium activity, generated intracellularly as a response to the fast synaptic activity, can modulate the spiking pattern of neural cells by changing their rate of firing and precisely synchronize them. This interaction of astrocytes and neurons could be used in generic NAN applications to add a new layer of information processing. One benefit of such periodic neural modulation is a predictable, logarithmic change in entropy rate (bits/spike) as a function of firing rate[89]. Interestingly, certain brain regions interact via elevations in their spike activity during a time window; a possible reason for this is that an increase of the firing rate decreases the number of bits per spike and, at least in the case of encoding an analog signal, reduces coding error[19]. By emulating the functional organization of NAN, as well as the dynamics of astrocytic Calcium activity and astrocyte-neuron interactions, we aim to computationally reveal the role of information processing on different temporal and spatial scales in brain networks. The broader theoretical benefits of having parallel processing on different temporal and spatial scales is an open question in computational brain science, one which will become increasingly important as our understanding of astrocytic intra- and inter-cellular Calcium dynamics evolves and expands.

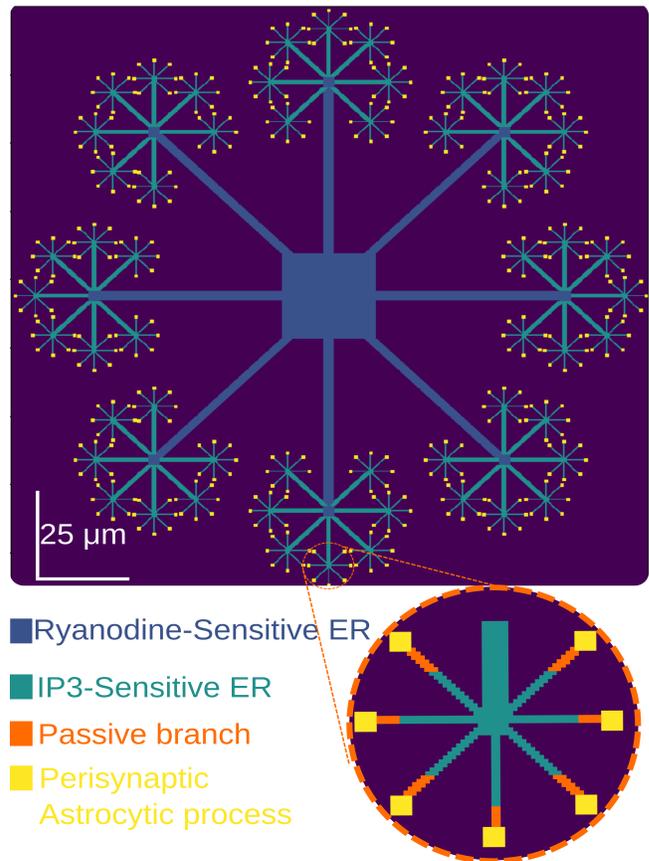

**Figure 1: Our proposed 2D model of an astrocyte controlling a spatial neural domain. The model preserves the basic compartments, their geometry and their main organelles.**

## METHODS

### Development of a Computational Model of an Astrocytic Cell

Astrocytes communicate with the neighboring neurons by ensheathing their synapses with thin, tendril-like processes[33]. A single astrocyte can form up to $10^5$ three body arrangements (astrocytic process, pre-synaptic neuron, post-synaptic neuron) known as tripartite synapses[3]. Through these structures, astrocytes are capable of sensing and modulating neuronal activity. Specifically, they are considered to encode synaptic activity into changes of their Calcium concentrations; three distinct spatial/temporal scales of astrocytic Calcium dynamics have been observed: slow intercellular waves[25, 59]; slow intracellular waves[25, 59]; and fast and slow localized transients in response to neural activity[6]. Then, they signal back to neurons in a Calcium-dependent manner by releasing gliotransmitters with both excitatory (e.g. glutamate) and inhibitory (e.g. gamma aminobutyric acid - GABA) effect. While these interactions between modes of Calcium response and the release of gliotransmitters, as well as the computational functions they subserve, are largely unknown, this bidirectional signaling may give rise to many interesting computational phenomena that we have recently started to study[46, 47].

Table 1: Dimensions of the Cell's Compartments

| Compartment / Dimension | Cell's Soma C6 | Thick branch C5 | Thin branch C4 | Fine branch C3 | Passive Fine branch C2 | Perisynaptic Process C1 |
|---|---|---|---|---|---|---|
| Length ($\mu m$) | 25 | 50 | 23 | 5 | 2 | 2.5 |
| Width ($\mu m$) | 25 | 3 | 1 | 0.2 | 0.2 | 2.5 |

Our proposed model preserves the experimentally reported basic astrocytic morphology and their associated functional compartments. Specifically, the cell consists of 6 different compartments, with increasing functional complexity going from the periphery to the center of the cell, as shown in Figure 1: (C1) The perisynaptic astrocytic processes with $IP_3$ dynamics; (C2) The passive part of the fine branches with $IP_3$ decay dynamics (orange); (C3) The fine branches with $IP_3$-Endoplasmatic Reticulum (ER) dynamics (teal); (C4) The thin branches with $IP_3$-ER dynamics (teal); (C5) The thick branches with Ryanodine-$IP_3$-ER dynamics (blue); (C6) The soma with Ryanodine-$IP_3$-ER dynamics (blue). The dimensions of the compartments are shown in Table 1.

The perisynaptic astrocytic processes (C1) and the 2 $\mu m$ long passive fine branch (C2) were devoid of ER[67, 75]. Therefore, at (C1), only $IP_3$ production and degradation[77] were modeled, as follows:

$$\dot{I} = I_\beta + I_\delta - I_{3K} - I_{5P}, \quad (1)$$

where I is the $IP_3$ concentration, $I_\beta$ and $I_\delta$ denote the agonist-dependent and the agonist-independent $IP_3$ production respectively, while the last two terms model the $IP_3$ degradation by two different enzymes, as described by De Pitta et al[18]. At (C2) no $IP_3$ is produced, but it can be degraded; therefore only the last two terms of Equation 1 are used.

Calcium dynamics were predominantly driven by the ER, which was absent in (C1) and (C2). This lack of ER, the main mechanism that regulates cytosolic Calcium concentration, would result in its accumulation through diffusion. The large set of molecular mechanisms (e.g. Sodium-Calcium exchanger, mitochondria) that contribute to the maintenance of Calcium concentrations within certain ranges was modeled in these compartments with the following Hill equation:

$$J_{bal} = -v_{bal}\frac{C^2}{C^2 + K_{bal}^2}, \quad (2)$$

where C denotes the intracellular Calcium concentration, $v_{bal} = 0.5\mu M/s$ is the maximal rate of Calcium depletion and $K_{bal} = 2\mu M$ is the Calcium affinity of the balancing mechanism.

The regulation of Calcium[49] in compartments (C3) and (C4) is described by Equations 3 and 4:

$$\dot{C} = J_{chan} + J_{leak} - J_{pump}, \quad (3)$$

where $J_{chan}$ denotes the increase of the cytosolic Calcium due to ER-release, $J_{leak}$ denotes the leakage from the ER and $J_{pump}$ denotes the absorption of the Calcium back into the ER by the SERCA pumps of the ER[18],

$$\dot{h} = \frac{h_\infty - h}{\tau_h} \quad (4)$$

where h is a gating variable, which controls the Calcium- and $IP_3$-dependent opening and closing of the channels on the ER surface that release Calcium.

Compartments (C5) and (C6), in addition to their $IP_3$-sensitive ER, they house the independent Ryanodine-sensitive ER[31, 32], which resides in the thick branches[41, 88]. This was added to Equation 3 to emulate its effects on Calcium dynamics:

$$J_{RyR} = \left(k_1 + k_2\frac{C^3}{C^3 + K_d^3}\right)(C_{ER} - C) \quad (5)$$

where $C_{ER}$ is the Calcium concentration in the ER. The parameters $k_1, k_2$, and $K_d$ above were fit to the experimental data by Bezprozvany et al[9] describing the dependence of Ryanodine-sensitive receptors on the Calcium concentration.

## Synaptic (Neural) Activity

The astrocytic model was stimulated using the output of the classical facilitation-depression model[95], which predicts the amount of neurotransmitter released by a presynaptic terminal. A fraction of this amount was received by the astrocyte[52] and fed into the term $I_\beta$ of Equation 1 driving the agonist-dependent $IP_3$ production in (C1). To close the NAN loop, we modeled the Calcium induced astrocytic gliotransmission to stimulate basally active postsynaptic neurons through the mechanism of Slow Inward Currents (SICs).

The postsynaptic neurons engulfed by the astrocytic compartment (C1) were modeled using the Leaky Integrate and Fire (LIF) model with a membrane resistance of $R = 0.6 G\Omega$ and a capacitance of $C = 100 pF$. The neurons fired at low, constant rate due to a basal current. A Slow Inward Current (SIC) was injected into the postsynaptic neuron at the peak of the astrocytic wave[65]. The amplitude dependence of SICs on astrocytic Calcium level followed the experimentally fit function[58, 63], described in Equations 6-8:

$$I_{astro} = 2.11\frac{\mu A}{cm^2}ln(w)\Theta(lnw) \quad (6)$$

where

$$w = [Ca^{2+}]/nM - 196.11 \quad (7)$$

and $\Theta(x)$ is the Heaviside function. The total equation for SIC dynamics, activated at every Calcium concentration peak was:

$$I_{SIC}(t) = I_{astro}([Ca_{peak}^{2+}])*n*\left(exp\left(\frac{t}{\tau_{decay}^{SIC}}\right) - exp\left(\frac{t}{\tau_{rise}^{SIC}}\right)\right) \quad (8)$$

where $\tau_{rise}^{SIC} = 50ms$ and $\tau_{decay}^{SIC} = 300ms$ are the time constants of the SIC and $n = \frac{6*6^{0.2}}{5}$ is a normalization constant[47] for exponential distributions, which assures that the biexponential distribution, before being multiplied by $I_{astro}$, is equal to one. Therefore the

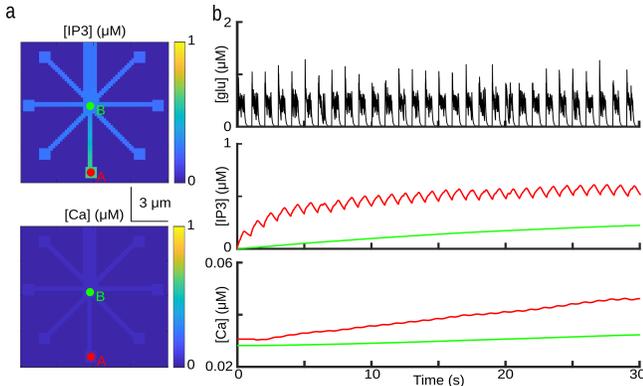

Figure 2: A constant activation of a single synapse releases enough glutamate (upper panel) to increase slowly the $IP_3$ production near the synapse (middle panel) but the amount of $IP_3$ produced by this stimulation is not sufficient to induce a Calcium wave (bottom panel).

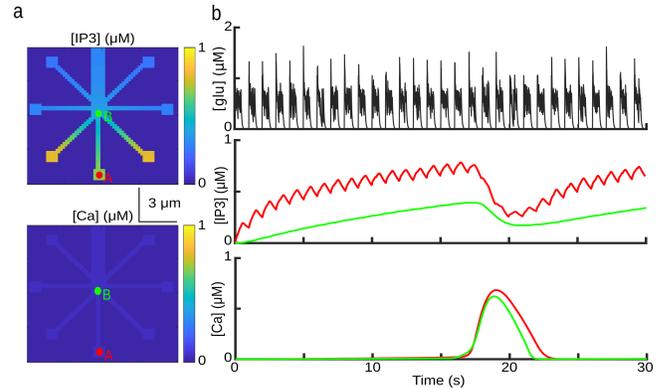

Figure 3: A simultaneous activation of three synapses releases enough glutamate (upper panel) to increase fast enough the $IP_3$ production near the synapse (middle panel) and induce a Calcium wave (bottom panel) that gets propagated towards the astrocytic soma.

resulting total current driving the postsynaptic neurons is modeled by Equation 9:

$$I_{total} = I_{basal} + I_{SIC} \quad (9)$$

where $I_{basal}$ is a basal presynaptic current.

## RESULTS

Here, we show how an astrocyte senses and encodes the extent of synchronous synaptic activity fusing the information from its microdomains into the amplitude and frequency of its intra-cellular Calcium wave. The wave gets generated on the peripheral thin branches of the cell and, if prominent enough, can reach its soma. We also show how a small local Calcium wave in the microdomain can synchronize neural firings. Strikingly, this result is in alignment with recent experimental findings that show how astrocytes can synchronize neurons at the same time scale (approximately 30 seconds) [23]. Our results are further described below.

First, the model suggested that the activity of a single synapse was enough to cause an increase in the $IP_3$ concentration, but this increase was not strong enough to generate a Calcium wave. We stimulated only one synapse in a microdomain of our model. The lack of activity in neighboring synapses prevents the accumulation of sufficient $IP_3$ concentration to induce a Calcium wave. This is shown in Figure 2. The discrepancy from the latest studies[10], which report astrocytes detecting activity even in a single synapse, might be due to the modelling of solely intracellular mechanisms. Another possible explanation could be that the stimulation pattern (1Hz bursts with 50Hz firing rate and 0.5s inter-burst interval for total stimulation 30s) was not sufficient to induce a Calcium wave.

Second, concurrent synaptic activity induced a Calcium wave. We synchronously activated a subset of synapses in the same microdomain. The concomitant activity in three of the seven synapses constrained the diffusion of $IP_3$, resulting in its increased concentration. The additive effect of the production by multiple synapses lead to $IP_3$ levels capable of generating a Calcium wave. This is shown in Figure 3. We should note that the wave reaches the branching point (point B) and the perisynaptic process (point A) almost simultaneously, although it is generated closer to the process. This happens due to the active mechanism of Calcium-Induced Calcium Release (CICR)[73] in compartments (C3) and (C4), unlike the passive diffusive process in compartments (C1) and (C2). Interestingly, this time lag between the initiation of the stimulus and the intracellular Calcium increase in our model aligns closely with the experimental results reported by Shigetomi et al [80], who also found a ~20 seconds delay between the stimulation of astrocytic glutamate receptors with PAR-1 agonists and the emergence of a Calcium increase. This group also reported that the Ca rises due to this kind of receptors' activation gave rise to SICs, while activation of $P2Y_1$ receptors did not.

Third, the Calcium wave synchronized the postsynaptic neurons of a microdomain. We used the aforementioned stimulation protocol of a subset of synapses to generate the $IP_3$-induced Calcium wave. The diffusion of the Calcium wave through the entire microdomain caused simultaneous release of gliotransmitters at the endfeet of compartment (C4). Transitively, these molecules induced SICs to the postsynaptic neurons, synchronizing their firing activity. This is shown in Figure 6. This bidirectional interaction between neurons and astrocytes suggests the notion of NANs as an interlaced functional unit by imposing synchronization on all synapses engulfed by a single microdomain.

Fourth, maximal synaptic engagement within a microdomain does not result in a global Calcium wave. We stimulated all seven synapses of the microdomain. This induced a Calcium wave in compartment (C4), as expected. However, upon entering compartment (C5) the wave had withered. Hence, it could not propagate towards the soma. This is shown in Figure 4. As in the case of one active synapse (see Figure 2), the activation of a single microdomain is marginalized by the lack of activity in its neighbors. This kind of Calcium activity could be interpreted as a medium-scale Calcium wave.

Lastly, synchronous activity in neighboring microdomains resulted in a global Calcium wave reaching the cell's soma (C6). This is shown in Figure 5. We sequentially stimulated microdomains with a time step of 1s. As expected, all microdomains generated Calcium waves. In spite of the temporal misalignment of the Calcium waves, their superposition resulted in a global wave. The convergence of two Calcium waves (orange and pink) pushed the cumulative concentration (yellow) at their collision point (point J) beyond the Ryanodine-sensitive ER activation threshold[9]. Then, a self-sustained Calcium wave was initiated and it propagated to the cell's soma (C6).

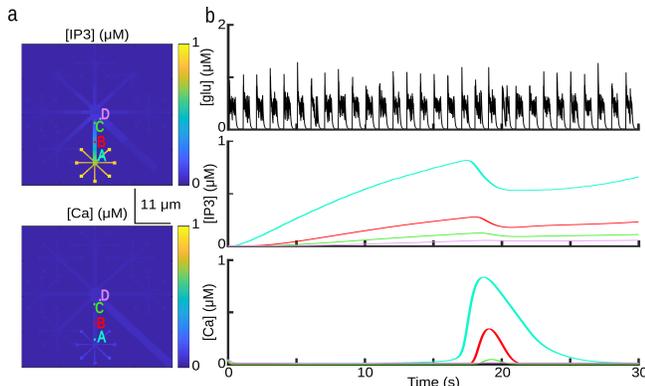

**Figure 4: The Calcium wave that was generated by the synchronous neural activity decreases in amplitude as it reaches the soma.** A simultaneous activation of all seven synapses releases enough glutamate (upper panel) to increase the $IP_3$ production near the synapse (middle panel) and induce a Calcium wave (bottom panel) that propagates towards the astrocytic soma. Close to its generation location (A), the Calcium wave has considerable amplitude which decreases as it propagates through points B and C of the intermediate branch until the branching point D. In point D, the impact that the Calcium wave may have on the thick branch is negligible. This can be interpreted as a medium-scale Calcium wave that is experimentally observed.

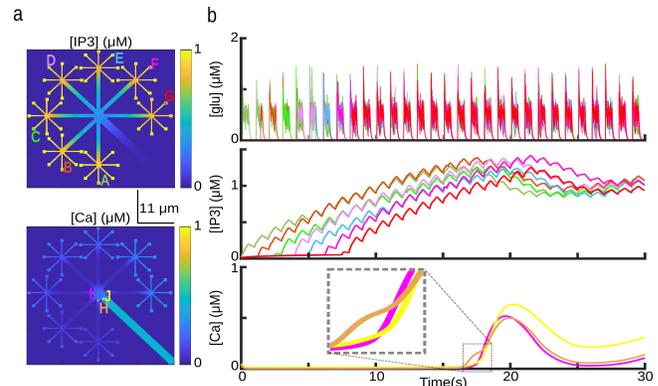

**Figure 5: a. Sequential stimulation of microdomains and its effect on the generated Calcium wave.** We stimulated all seven synapses on each microdomain, so that each microdomain is stimulated sequentially. In other words, point A was stimulated first, followed by points B,C,D,E,F and G, at this order. The timing of glutamate release in the microdomains is shown(b , upper). The serial stimulation induces a serial increase of $IP_3$ in the microdomains (b, middle). Slight geometrical differences among the dimensions of the perisynaptic processes cause the non-homogeneous increase of the $IP_3$ concentration in each of them, as in real cells. The fluctuation of the Calcium concentration at H, I and J is shown (b, bottom). At the stem(H) of the first stimulated microdomain, there was an initial increase of Calcium concentration. By the time this increase got saturated, the Calcium wave(I) from the microdomain stimulated second arrived and was superimposed to the existent amount. The summed Calcium concentration was sufficient to trigger the global Calcium wave, which propagated along the thick branch. This is pointed out in the zoomed-in area. The increase in the slope of the concentration at point J happens, when the two waves at points H and I merge.

## DISCUSSION

In this paper, we presented a biologically plausible 2D astrocytic model that preserved the geometrical structure as well as the main compartments and their associated organelles found in its biological counterpart. Stepping away from modeling astrocytes as point processes, we showed how preserving the geometrical and functional characteristics of these cells might elucidate how their Calcium waves are shaped and how their spatial extents may encode, and modulate, the nearby synaptic activity, at least in terms of synchronizing neurons. The novelty of our approach lies not only in the structural and functional compartmentalization of an astrocyte that aligns with recent experimental results[31–33, 41, 67, 75, 77, 88], but also in the functional role that these spatially distinct intracellular mechanisms have, endowing the astrocyte with a mechanism to sense the synaptic activity and impose, even preserve, neural synchronization. This is crucial for understanding how biological intelligence emerges from neural activity, because oscillations, i.e., synchronous neural activity, are believed to convey information in space and time[79, 87] by forming neural networks, via local and longer range synchronization[5, 27, 85]. A previous study on astrocyte-induced neuronal synchronization is proposed for neuromorphic circuits, with promising results. [42].

Our model suggests that the activity of neighboring synapses may be integrated by astrocytes and encoded into Calcium waves of different extents, reaching different parts of the cell. Interestingly, such compartmentalization is verified experimentally by studies employing genetically encoded Calcium indicators[12, 39, 94] showing that astrocytes display diverse Calcium signals within sub-compartments[36, 81, 82, 84, 86]. When stimulation patterns within the biological range were also applied to our astrocytic model, we observed 1) medium-scale Calcium waves that propagated towards the thick branches and 2) full-scale "global" Calcium waves that propagated along the thick branches to the cell's soma. These results align with recent experimental findings that report both types of Calcium waves [20, 33, 44, 83, 98]. Strikingly, none of these studies were able to identify possible generative mechanisms or functional roles for the observed waves. Our model provides such possible explanations and suggests how the synaptic-induced Calcium waves, through gliotransmission[63], can influence synaptic activity in silent neurons which are controlled by the active astrocyte, depending on the extent of the generated Calcium waves. This could have further implications in understanding both biological and artificial networks, given that gliotransmission has been linked to cortical state switching[72], learning and neuronal synchronization[2, 23, 43, 72, 92].

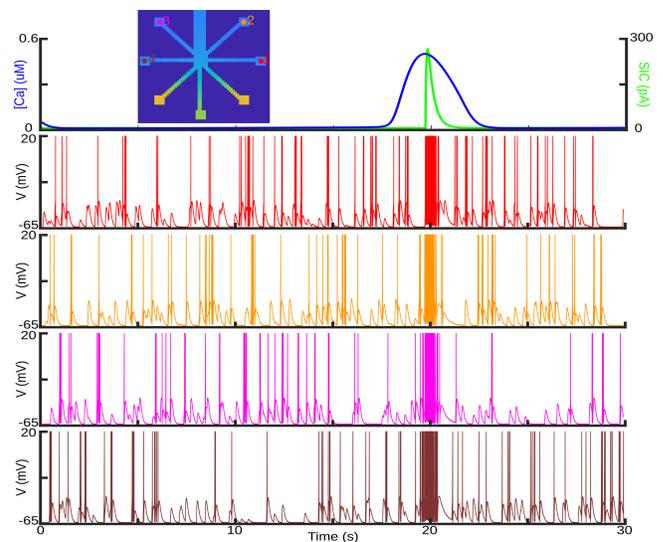

**Figure 6: A local Calcium wave synchronizes neurons controlled by an astrocytic microdomain.** The concurrent activity of 3, out of 7, synapses causes a local Calcium wave that reaches the non-active neurons and synchronizes them (at 20 seconds).

Surprisingly, rather sporadic efforts for translating the astrocytic mechanisms into AI currently exist, mainly focused on improving the classification performance of classical neural networks, through synaptic potentiation[1, 71]. A quite recent, second, approach involves astrocytes into co-evolutionary learning, at the benefit of avoiding the manual parameter tuning for each specific problem[56]. In 2016, NANs have been identified as potential targets for deep learning[66]. The overlay of the astrocytic layer on the neuronal network layer could also possibly improve the function of our brain-morphic robots, which exploit the neuronal connectome [90].

Our proposed astrocytic model suggests a functional role for Calcium waves in NAN applications, where astrocytes may be sensing and imposing synchronous neural activity across their microdomains. Incorporating astrocytes into conventional neural networks shows a great potential to increase their computational capabilities. Take for example the tiling organization of astrocytes, which is preserved across brain areas and species[13, 64, 99]. Since astrocytic networks seem to facilitate diffusion by forming a discrete lattice, a fascinating possibility is that they might routinely carry out diffusive computations. In other words, the diffusive properties of Calcium waves might naturally encode a smooth version of neural representations. Solving such problems for the simulated astrocytes would enable the formulation of testable hypotheses on how their biological counterparts may fuse diffused Calcium signals to sense information[34] sent from their end-feet and, in response, release global Calcium waves, neuronally-mediated[21, 36] or reported as "spontaneous"[6, 38, 48, 64], modulating other astrocytes and neurons.

A future direction of the model would include the embodiment of other intracellular mechanisms, which for now have been omitted, to reproduce the most recently observed highly localized Calcium waves, which are believed to take place in much shorter time scales (milliseconds). Overall, further exploring how local, medium and global astrocytic signals contribute to the functioning of neural circuits is a direction worth pursuing.

## Conclusion

Prevailing over a century, the neuronal paradigm of studying the brain has left limitations in both our understanding of how the brain processes information to achieve biological intelligence and how such knowledge is translated into artificial intelligence. Overturning our assumptions of how the brain functions, the recent exploration of astrocytes, the most abundant yet long-neglected non-neuronal brain cells, has ignited a revolution in our fundamental understanding of biological intelligence. Our proposed biologically plausible astrocytic model, on one hand, is helping us study how astrocytes work independently but cooperatively with the neuronal brain and, on the other hand, paves the way for developing astrocyte-derived computational principles for artificial intelligence. Translating the mechanisms that astrocytes have to process information and modulate the neuronal activity into brain-morphic computing is an exciting research area that we are actively exploring. It has not escaped our attention that such efforts may spur brain-derived non-von Neumann architectures and vitalize intelligent cognitive assistants.